\documentclass[12pt]{article}
\title{
\vspace{-3mm}
\rightline{\small HU-EP-05/20}
\vspace{8mm}
\bf Influence of matter fields on the (de-)confining properties of the 3d Georgi--Glashow model}
\author{Dmitri Antonov \thanks{Permanent address: ITEP, 
B. Cheremushkinskaya 25, RU-117 218 Moscow, Russia.}\\
{\it Institute of Physics, Humboldt University of Berlin,}\\
{\it Newtonstr. 15, 12489 Berlin, Germany}}
\date{}
\begin{document}
\maketitle
\vspace{1mm}
\centerline{\bf {Abstract}}
\vspace{3mm}
\noindent
The influence of various matter fields on the confining and finite-temperature properties of the (2+1)d 
Georgi--Glashow model is explored.
At zero temperature, these fields are W-bosons, which play the role of heavy nodes, through which the quark--anti-quark string 
passes. This fact is shown to increase by a factor $4\sqrt{2}$ the absolute value of the coefficient at the 
$1/R$-term in the large-distance potential with respect to that of the Nambu--Goto string in 3d.
The string tension also acquires a positive correction, which is, however, exponentially small.\\
At finite temperature, the matter fields of interest are massless fundamental quarks, which diminish the deconfinement
critical temperature by way of an additional attraction of a monopole and an anti-monopole inside their molecules 
through the quark zero modes. It is demonstrated that, outside the BPS-limit, when the number of massless
flavors is 4 or larger, the deconfinement phase transition occurs already at the temperatures
of the order of the temperature of dimensional reduction. In the BPS limit, this critical number of flavors is 3.
Since the temperature of dimensional reduction is exponenetially small and since monopoles
are instantons in (2+1)d, these numbers can be compared with the one
in the instanton-liquid model of 4d QCD, 
at which the chiral phase transition occurs at a vanishingly small temperature. The latter number is known to be of the order of 5, so that the 
results of the two models are quite close to each other.

\vspace{6mm}
\noindent

\section{Introduction; the (2+1)d Georgi--Glashow model}
The (2+1)d Georgi--Glashow (GG) model is a classic example of a gauge theory where confinement can be studied 
analytically~\cite{pol}. The advantage of this model over QCD is that it supports confinement in the 
weak-coupling regime, where the vacuum of the GG model is already nonperturbative, since it is 
populated by 't~Hooft--Polyakov monopoles~\cite{mon}. The latter form a plasma, which, with an
exponentially high accuracy, is dilute and consists of monopoles and anti-monopoles of a unit magnetic charge.
Random magnetic fluxes in this plasma through the closed trajectory of an external quark--anti-quark
pair produce then confinement of external fundamental quarks. In another words, in the monopole--anti-monopole plasma, 
a dual photon becomes Debye screened and develops by this mechanism an exponentially small, but finite, mass.
Accordingly, the vacuum correlation length becomes finite (rather than infinite, as without this mechanism),
and a string is formed between a quark and an anti-quark when they are separated by the distances larger than 
this length.

The Euclidean action of the GG model reads
$$
S=\int
d^3x\left[\frac{1}{4g^2}\left(F_{\mu\nu}^a\right)^2+\frac12
\left(D_\mu\Phi^a\right)^2+ 
\frac{\lambda}{4}\left(\left(\Phi^a\right)^2-\eta^2\right)^2\right],
$$
where the Higgs field $\Phi^a$ transforms by the adjoint
representation, i.e.
$D_\mu\Phi^a\equiv\partial_\mu\Phi^a+\varepsilon^{abc}A_\mu^b\Phi^c$.
The weak-coupling regime $g^2\ll M_W$, which will be assumed henceforth, parallels the requirement
that $\eta$ should be large enough to ensure spontaneous symmetry
breaking from SU(2) to U(1). At the perturbative level, the spectrum
of the model in the Higgs phase consists of a massless photon, two  
charged W-bosons with mass $M_W=g\eta$, and a neutral Higgs
field with mass $M_H=\eta\sqrt{2\lambda}$.

The 't~Hooft--Polyakov monopole, which represents the nonperturbative contents of the model, is a solution to the
classical equations of motion with the following Higgs- and
vector-field parts:
\vspace{-2mm}
\begin{equation}
\label{scal}
\Phi^a=\delta^{a3}u(r),~~ u(0)=0,~~ 
u(r)\stackrel{r\to\infty}{\longrightarrow}\eta-\exp(-M_Hr)/(gr);
\end{equation}
\vspace{-2mm}
\begin{equation}
\label{vect}
A_\mu^{1,2}(\vec x)\stackrel{r\to\infty}{\longrightarrow}
{\mathcal O}\left({\rm e}^{-M_Wr}\right),~~
H_\mu\equiv\varepsilon_{\mu\nu\lambda}\partial_\nu A_\lambda^3=
\frac{x_\mu}{r^3}-4\pi\delta(x_1)\delta(x_2)
\theta(x_3)\delta_{\mu 3};
\end{equation}
as well as the following action $S_0=4\pi M_W\epsilon/g^2$. Here,
$\epsilon=\epsilon\left(M_H/M_W\right)$ is a certain monotonic, slowly
varying function, $\epsilon\ge 1$, $\epsilon(0)=1$ (BPS-limit)~\cite{bps},
$\epsilon(\infty)\simeq 1.787$~\cite{kirk}.
The statistical weight of a single monopole in the grand canonical ensemble (the so-called fugacity)
has the form~\cite{pol} $\zeta=\delta(M_H/M_W)\frac{M_W^{7/2}}{g}{\rm e}^{-S_0}$.
The function $\delta$ here is known~\cite{ks} to grow in
the vicinity of the origin (i.e. in the BPS limit), but the
speed of this growth is such that it does not spoil the exponential
smallness of $\zeta$. As follows from Eq.~(\ref{vect}), the interaction of monopoles through the vector part of the 
't~Hooft--Polyakov solution is determined by the Coulomb force, whose strength is proportional to the square of the magnetic coupling constant $g_m=4\pi/g$.
In the so-called compact-QED limit, where $M_H\ge{\cal O}(M_W)$,
the summation over the grand canonical ensemble of monopoles leads to the 
action $S=\int d^3x\left[\frac12(\partial_\mu\chi)^2-2\zeta\cos(g_m\chi)\right]$, where 
$\chi$ is the dual-photon field. The mean monopole density stemming from this expression, $2\zeta$,
yields the mean distance between nearest neighbors in the plasma of the order of $\zeta^{-1/3}$.
This distance is exponentially larger than the monopole size, $M_W^{-1}$, that justifies 
the dilute-plasma approximation. The Debye mass of the dual photon, stemming from the expansion of the cosine, reads
$M_D=g_m\sqrt{2\zeta}$. It is therefore exponentially smaller than the next (in the order of largeness) natural 
mass parameter of the theory, $g^2$ (which itself is much smaller than $M_W$). Furthermore, one can prove that
the number of monopoles contained in the Debye volume, $M_D^{-3}$, is exponentially large, namely 
$2\zeta M_D^{-3}=\frac{1}{g_m^3\sqrt{2\zeta}}\gg 1$. This justifies the validity of the mean-field approximation,
which implies that the fluctuations of fields of individual monopoles can be disregarded, and the  
grand canonical ensemble can be described in terms of only one, dual-photon, field.

Since the pioneering Polyakov's paper~\cite{pol}, it is argued that
W-bosons do not affect the infra-red properties of the theory because of their heaviness.
In the next Section, we will make this statement quantitative, by proving that the relative correction to the 
tension of the quark--anti-quark string, produced by W-bosons, is exponentially small. On the other hand, it will be 
shown that W-bosons increase the absolute value of the coefficient of the (next-to-linear) $1/R$-term in the quark--anti-quark potential 
(the so-called L\"uscher term~\cite{55, lusch})
from that of the bosonic-string theory (which is $\pi/24$ in 3d) to $\pi/(3\sqrt{2})$.
This effect takes place for long enough quark--anti-quark strings, namely those whose length well exceeds the 
maximal distance between two W-bosons, at which the adjoint string between them still exist. [Such a distance, at which
adjoint strings break due to the production of W$^\pm$-pairs, is parametrically ${\cal O}\left(\frac{M_W}{g^2M_D}\right)$.] 
This is the effect of matter fields on the {\it confining} 
properties of the GG model, we consider in the present note.

At finite temperatures, W-bosons play the crucial role on the deconfinement phase transition, as was for the first time
realized in~\cite{gg1} (see~\cite{ggrev, dc} for reviews). Rather, the subject which is yet not fully understood
is the influence of {\it external} matter fields on the finite-temperature properties of the 
model. Some partial progress in this direction has been achieved in~\cite{q, dkn}.~\footnote{Note also Ref.~\cite{ak},
where the additional matter field was the superpartner of the dual photon -- the two-component Majorana spinor.}
In both cases, external matter was 
transforming by the fundamental representation. In Ref.~\cite{dkn}, the matter 
fields were represented by heavy scalar bosons, whereas in~\cite{q} -- 
by massless spin-$\frac12$ quarks. The latter analysis has been performed not in the GG model but in the (continuum limit of) 3d compact QED,
where W-bosons were absent. In Section~3, we will consider the full GG model with massless fundamental quarks included.
We will see that, just as in the absence of W-bosons, the deconfinement critical temperature becomes reduced by massless quarks 
by a factor of the order of 1.~\footnote{Instead, when quarks are massive, the deconfinement critical temperature reduces by an 
exponentially large factor and becomes as small as the temperature of dimensional reduction~\cite{q}.} The reason for this reduction 
is an additional attraction of a monopole and an anti-monopole inside a molecule (above the deconfinement critical temperature),
which is produced by the quark zero modes in the monopole field. We will find the number 
of massless quark flavors at which the deconfinement phase transition occurs already at the temperatures which are as exponentially small
as the temperature of dimensional reduction in our model. 
The main results of the Letter will be summarized in Section~4. Finally, some technical details will be presented in Appendix.

\section{Influence of W-bosons on the quark--anti-quark potential}
The fundamental-string tension in the GG model reads~\cite{pol} (see~\cite{dc} for a review) $\sigma=\frac{c}{4\pi}g^2M_D$. 
Since the thickness of the string, equal to the vacuum correlation length $M_D^{-1}$, 
is exponentially large (even with respect to the mean distance
between monopoles in the plasma, which is ${\cal O}(\zeta^{-1/3})$),
the value of the 
coefficient $c$ here depends on the range of averaging the magnetic field (whose flux through the contour of the 
Wilson loop produces confinement) in the direction perpendicular to the world sheet. For instance, for a flat world sheet 
(i.e. a straight-line string), if one chooses the value of the magnetic field right on the world sheet, 
then $c=4$.~\footnote{For comparison, the value of $c$
one obtains in case when the density of the monopole plasma is much lower than the mean one, $2\zeta$, is
$\pi$. Remarkably, this 
is true for an arbitrarily-shaped world sheet and not only for a flat one. For a flat world sheet, 
this value of $c$ can be shown to approximately correspond to the range of averaging the magnetic field  
$|z|<\frac{\sqrt{6-\frac{3\pi}{2}}}{M_D}$, where $z$ is the coordinate transverse to the world sheet.}
In the absence of external matter fields, W-bosons exist in the molecular phase, which means that a ${\rm W}^{+}$- and a ${\rm W}^{-}$-bosons
are bound into a molecule, where they interact through an adjoint string. The length of this string (i.e. the size of a molecule)
may vary from its thickness, $l_{\rm min}=M_D^{-1}$, to the distance at which the energy of the string is enough to produce a 
${\rm W}^{+}-{\rm W}^{-}$--pair
out of the vacuum. This distance therefore reads $l_{\rm max}=\frac{2M_W}{\sigma}=\frac{8\pi M_W}{cg^2M_D}$, that is larger than
$l_{\rm min}$ by a big factor of the order of $M_W/g^2$. When an infinitely heavy quark and an anti-quark are inserted into the 
system and separated by asymptotically large distances of interest, the corresponding fundamental string starts passing through 
those W-bosons, which it meets in between.~\footnote{The corresponding ${\rm W}^{+}-{\rm W}^{-}$--molecules recombine in the sense that one 
of the two fundamental strings, which form the adjoint string of a molecule, flips onto a W-boson belonging to another molecule and so on.}
It is easy to imagine the structure of such a piecewise string from the requirement that its energy tends to be the minimal possible one.
Firstly, this leads to the conclusion that only W-bosons, which are    
the nearest ones to the line joining a quark and an anti-quark, are involved in this chain-like structure. Secondly,
it is energetically favorable to have a chain with the maximal possible distances between the neighbors, $l_{\rm max}$. Indeed,
let us consider a chain carrying $N$ W-bosons, so that $N+1=\frac{R}{l}$, where $l$ is the distance between some two nearest neighbors.
For the case $N\gg 1$ of interest, the energy of the chain reads $E=\sigma R+NM_W\simeq\left(\sigma+\frac{M_W}{l}\right)R$.
Therefore, the energy minimizes at $l=l_{\rm max}$, that proves the statement. Henceforth, in order to simplify the 
notations, we will denote the distance between the nearest neighbors in the chain as just $l$ and not $l_{\rm max}$.
Placing then the quark--anti-quark pair along the $x$-axis,
with a quark located at the origin and an anti-quark separated by the distance $R$, 
we can depict the chain as follows:~\footnote{In reality, the segments of the string are almost parallel to the $x$-axis.}

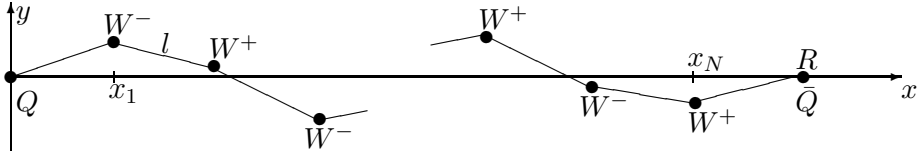
\begin{picture}(380,100)
\put(30,50){$\bullet$}
\put(35,40){{\small $Q$}}
\put(33,53){\vector(1,0){337}}
\put(33,25){\vector(0,1){56}}
\put(35,75){{\small $y$}}
\put(370,45){{\small $x$}}
\put(330,50){$\bullet$}
\put(330,40){{\small $\bar Q$}}
\put(33,53){\line(3,1){40}}
\put(69,63){$\bullet$}
\put(71,66){\line(4,-1){40}}
\put(107,54){$\bullet$}
\put(110,56){\line(2,-1){40}}
\put(147,34){$\bullet$}
\put(148,36){\line(5,1){20}}
\put(330,53){\line(-4,-1){40}}
\put(289,40){$\bullet$}
\put(291,43){\line(-6,1){40}}
\put(250,46){$\bullet$}
\put(252,49){\line(-2,1){40}}
\put(210,65){$\bullet$}
\put(212,69){\line(-5,-1){20}}
\put(90,62){{\small $l$}}
\put(72,55){\line(0,-1){4}}
\put(70,45){$x_1$}
\put(68,69){{\small $W^{-}$}}
\put(108,60){{\small $W^{+}$}}
\put(144,26){{\small $W^{-}$}}
\put(289,32){{\small $W^{+}$}}
\put(248,38){{\small $W^{-}$}}
\put(210,71){{\small $W^{+}$}}
\put(330,56){{\small $R$}}
\put(291,55){\line(0,-1){4}}
\put(289,57){{\small $x_N$}}
\end{picture}

\noindent
Thirdly, one can readily see that the potential 
between the nearest neighbors in the chain is linear with the accuracy, which is even higher than the exponential one.
Indeed, the potential between any two nearest neighbors in the chain is the one of the Nambu--Goto string~\cite{alv} 
$V_{{\rm W}^{+}{\rm W}^{-}}(r)=\sigma\sqrt{r^2-r_c^2}=\sigma r\left[1-\frac12(r_c/r)^2+{\cal O}\left((r_c/r)^4\right)\right]$,
where, in 3d,~\cite{lusch} $r_c=\sqrt{\frac{\pi}{12\sigma}}$. At $r=l_{\rm max}$, one has for the ratio of the leading term to the absolute
value of the next-to-leading one: $\frac{\sigma l_{\rm max}}{\pi/(24 l_{\rm max})}=\frac{384M_W}{cg^2}\frac{M_W}{M_D}$.  
This expression is larger than the exponentially large parameter $M_W/M_D$ by another very big factor $384M_W/(cg^2)$.
The interaction between any two neighbors is therefore linear with a very high accuracy.~\footnote{As will be shown below, the $1/R$-term in the 
quark--anti-quark potentail re-appears, albeit with a coefficient different from that 
of the Nambu--Goto string.}
Finally, taking into account that W-bosons are non-relativistic, we can write the Hamiltonian of the chain as 
$H=\sum\limits_{n=0}^{N}\left(\frac{{\bf p}_n^2}{2M_W}
+\sigma|{\bf r}_{n+1}-{\bf r}_n|\right)$, where ${\bf r}_0=0$, ${\bf r}_{N+1}=(R,0)$, ${\bf p}_0\equiv 0$.

The spectrum of this Hamiltonian can be found by noticing that, again due to their heaviness, W-bosons 
only slightly oscillate around their mean positions, i.e. $x_n\simeq nl+\xi_n$, where $|\Delta\xi_n|\equiv|\xi_{n+1}-\xi_n|\ll l$,
as well as $|\Delta y_n|\equiv|y_{n+1}-y_n|\ll l$. Therefore,
$|{\bf r}_{n+1}-{\bf r}_n|=\left[(l+\Delta\xi_n)^2+(\Delta y_n)^2\right]^{1/2}\simeq
l\left[1+\frac{\Delta\xi_n}{l}+\frac{(\Delta y_n)^2}{2l^2}\right]$.
Notice that the quadratic dependence on $\Delta\xi_n$ drops out from this expression.
The same happens to the linear dependence on $\Delta\xi_n$, since 
$\sum\limits_{n=0}^{N}\Delta\xi_n=\xi_{N+1}-\xi_0=0$.
Therefore, the motion along the $x$-axis is free and affects the 
spectrum only in the form of the constant $\sigma R$. Namely, we obtain
$\sigma\sum\limits_{n=0}^{N}|{\bf r}_{n+1}-{\bf r}_n|\simeq\sigma R+\frac{K}{2}\sum\limits_{n=0}^{N}(\Delta y_n)^2$,
where $K\equiv\sigma/l$. The problem of finding 
corrections to the large-distance linear quark-antiquark potential is, thus, reduced to the problem of
finding the spectrum of  
the following Hamiltonian, which describes transverse fluctuations:
$H=\sum\limits_{n=0}^{N}\left[\frac{p_n^{y{\,}2}}{2M_W}+\frac{K}{2}(\Delta y_n)^2\right]$.
For the case $N\gg1$ under study, this is a standard solid-state physics problem,
whose solution is presented in Appendix. The resulting quark-antiquark potential reads

\begin{equation}
\label{spectrum}
E(R)=\sigma R+2\sqrt{\frac{\sigma}{M_Wl}}\left[\cot\left(\frac{\pi l}{4R}\right)-1\right]=
(\sigma+\Delta\sigma)R-\frac{\alpha}{R}-2\sqrt{\frac{\sigma}{M_Wl}}
+{\cal O}\left(\frac{\sqrt{\sigma l^5/M_W}}{R^3}\right),
\end{equation}
where 

\begin{equation}
\label{coeffs}
\frac{\Delta\sigma}{\sigma}=\frac{8}{\pi}\frac{1}{\sqrt{\sigma M_W l^3}}=
\frac{c}{\sqrt{2}\pi^2}\frac{g^2M_D}{M_W^2},~~ 
\alpha=\frac{\pi}{6}\sqrt{\frac{\sigma l}{M_W}}=\frac{\pi}{3\sqrt{2}}.
\end{equation}
We see that the relative correction to the string tension is by a factor ${\cal O}(g^2/M_W)$ smaller than the 
exponentially small ratio $M_D/M_W$. However, 
W-bosons are quite not irrelevant to the infra-red properties of the GG model as far as the value of $\alpha$ is concerned.
Indeed, it is by a factor $4\sqrt{2}$ larger than that of the Nambu--Goto string in 3d, equal to $\pi/24$.

\section{Inclusion of fundamental quarks at finite temperature}
Dimensional reduction in the GG model is associated with the change of the 3d Coulomb interaction between monopoles
by the 2d one and occurs therefore at the temperatures ${\cal O}(\zeta^{1/3})$~\cite{gg}.
After the dimensional reduction, the model is described by the action~\cite{gg1}

\begin{equation}
\label{action}
S=\int d^2x\left[\frac12(\partial_\mu\chi)^2-2\xi\cos(g_m\sqrt{T}\chi)-2\mu\cos\tilde\chi\right].
\end{equation}
In this 2d theory, W-bosons are nothing but vortices of the dual photon, $\chi$, and their field, $\tilde\chi$, 
is defined through the relation 
$i\partial_\mu\tilde\chi=g\sqrt{\beta}\varepsilon_{\mu\nu}\partial_\nu\chi$. Here $\beta=1/T$,
$\xi=\beta\zeta$, and $\mu$ is the fugacity of W-bosons,
semiclassically equal to a half of their density, $\rho_W/2$. This density itself reads

$$
\rho_W=6\int\frac{d^2p}{(2\pi)^2}\frac{1}{\exp\left[\beta\left(M_W+\frac{p^2}{2M_W}\right)\right]-1}=
-\frac{3M_WT}{\pi}\ln\left(1-{\rm e}^{-\beta M_W}\right)\simeq\frac{3M_WT}{\pi}{\rm e}^{-\beta M_W},
$$
where the factor ``6'' in the initial expression describes the total number of spin states of W$^{+}$- and W$^{-}$-bosons.
In the last equality, we have used the fact that the temperatures of our interest (at which the deconfining 
phase transition occurs) are ${\cal O}(g^2)$, that is much smaller than $M_W$.~\footnote{On the other hand,
these temperatures are exponentially larger than the temperature of dimensional reduction, which means 
that the deconfining phase transition occurs deeply in the region where the theory is two-dimensional.} 
When one crosses the 
deconfining temperature, $T_c$, in the direction from smaller to larger temperatures, W-bosons become relevant
(and strings between them melt), while 
monopoles become irrelevant (and bind into molecules). 

In Ref.~\cite{gg1} (see~\cite{ggrev} for a review), two independent criteria for the determination of $T_c$ have been proposed.
One criterium is that $T_c$ is defined as a point where 
the densities of monopoles and W-bosons are equal, i.e. $2\xi=\rho_W$. Intuitively, it stems from the 
argument that the phase transition occurs when the thickness of the string becomes equal to its length.
While the thickness of the string is $M_D^{-1}\propto\zeta^{-1/2}$, its length is of the order of the 
mean distance between W-bosons, that is $\rho_W^{-1/2}$. Up to pre-exponential factors, the thickness 
and the length of the string are therefore equal at $T=g^2/(4\pi\epsilon)$. These qualitative 
arguments have further been supported by the RG approach~\cite{gg1}. Specifically, in the theory~(\ref{action}), the 
RG equations possess three fixed points. Among these, two are the zero- and the infinite-temperature ones, whereas 
the third fixed point is a non-trivial infra-red unstable one, $\xi=\mu$, $T=g^2/(4\pi)$, which should correspond
to the phase transition.

Another criterium (seemingly independent of the first one) is that $T_c$ is defined as a temperature, at which the
scaling dimensions $\Delta$ and $\tilde\Delta$ 
of the operators $:\!\cos\big(g_m\sqrt{T}\chi\big)\!:$ and $:\!\cos\tilde\chi\!:$ (in the theories 
where either only monopoles or W-bosons are present, respectively) coincide.
These scaling dimensions are equal $\frac{g_m^2T}{4\pi}$ and $\frac{g^2\beta}{4\pi}$
respectively, so that the monopole cosine term is relevant at $T<g^2/(2\pi)$~\cite{gg}, 
whereas the cosine term of the W-bosons 
is relevant at $T>g^2/(8\pi)$. The two scaling dimensions become equal when the following nice relation holds: $g_m^2T=g^2\beta$. 
This yields the critical temperature 
$g^2/(4\pi)$.~\footnote{At $T=g^2/(4\pi)$, both scaling dimensions are equal to unity,
therefore both cosine terms are relevant at this temperature. In fact, in the whole 
region of temperatures $g^2/(8\pi)<T<g^2/(2\pi)$, both terms are relevant.}

The fact that the above-described different criteria yielded so close values of the critical temperature 
(which could at most differ by a factor $\epsilon(\infty)\simeq 1.787$)
was remaining almost mysterious until the appearance of the paper~\cite{sk}. The authors of that paper have managed to overcome the 
problem existing in the RG approach of Ref.~\cite{gg1}. The essence of this problem is that, in the infra-red unstable 
fixed point, which defines the critical temperature, the two fugacities, $\xi$ and $\mu$, become not only equal to
each other, but also infinite, that apparently contradicts the dilute-plasma approximation. The authors of Ref.~\cite{sk}
have argued that the theory can only be critical if, at the same energy scale 
$\Lambda$, both $\xi(\Lambda)/\Lambda^2$ and $\mu(\Lambda)/\Lambda^2$ are of the order of 1.
Using the RG equations~\cite{j} (valid as long as both $\xi(\Lambda)/\Lambda^2$ and $\mu(\Lambda)/\Lambda^2$ do not exceed 1)
$$\frac{\partial}{\partial\lambda}\frac{\xi(\Lambda)}{\Lambda^2}=(2-\Delta)\frac{\xi(\Lambda)}{\Lambda^2},~~  
\frac{\partial}{\partial\lambda}\frac{\mu(\Lambda)}{\Lambda^2}=(2-\tilde\Delta)\frac{\mu(\Lambda)}{\Lambda^2},$$
where $\lambda=\ln(T/\Lambda)$, one has 
$$\frac{\xi(\Lambda)}{\Lambda^2}=\frac{\zeta}{T^3}\left(\frac{T}{\Lambda}\right)^{2-\Delta},~~
\frac{\mu(\Lambda)}{\Lambda^2}=\frac{\mu}{T^2}\left(\frac{T}{\Lambda}\right)^{2-\tilde\Delta}.$$ 
Equating these expressions to 1, one then has 
(disregarding the inessential pre-exponential factors): 
$-S_0+(2-\Delta)\lambda=-\beta M_W+(2-\tilde\Delta)\lambda=0$. Notice that 
this equation contains the information on the fugacities {\it and} on the scaling dimensions. 
The corresponding critical temperature is~\cite{sk} $T_c=\frac{g^2}{4\pi}\frac{2+\epsilon}{1+2\epsilon}$. It
reproduces correctly both the compact-QED result~\cite{gg} at $\epsilon\to 0$ (when the density of monopoles is 
exponentially larger than the density of W-bosons) and the Berezinsky--Kosterlitz--Thouless 
critical temperature in the pure 2d plasma of W-bosons
in the opposite limit $\epsilon\to\infty$.~\footnote{Clearly, both limits are formal and can never be physically 
realized in the GG model.} 

In Ref.~\cite{q}, it has been shown that, when $N_f$ flavors of 
massless dynamical fundamental quarks are introduced into 3d compact QED,
the scaling dimension of monopoles changes as $\Delta\to\Delta_{N_f}=\Delta+N_f$.
That is because quark zero modes in the monopole field produce an additional attraction 
of a monopole and an anti-monopole inside a molecule.~\footnote{In another words, 
the action of a molecule in the presence of $N_f$ massless fundamental flavors 
is modified as $S_{M\bar M}=\frac{g_m^2T+4\pi N_f}{2\pi}\ln(\tilde\mu R)$, where $\tilde\mu$ is the infra-red cutoff.} 
For this reason, the critical 
number of massless fundamental flavors in 3d compact QED 
is just 2. This means that, at $N_f\ge 2$, the phase transition from the 
phase of the monopole plasma to the phase of monopole--anti-monopole molecules occurs at the same temperatures as the 
dimensional reduction, i.e. ${\cal O}(\zeta^{1/3})$. In this sense we will imply the notion of the  
critical number of flavors also below.

Let us now consider the full GG model, rather than just 3d compact QED, and extend it by fundamental quarks.
One can see that the change of $\Delta$ leads then to two quite different 
results -- one stems from the condition $\Delta_{N_f}=\tilde\Delta$, and the other, correct one, -- from the condition 
$\frac{\xi(\Lambda)}{\Lambda^2}=\frac{\mu(\Lambda)}{\Lambda^2}\sim 1$. Indeed, in the first case, the critical temperature is determined from the 
equation $g_m^2T+4\pi N_f=g^2\beta$.~\footnote{With the notation $x=4\pi T/g^2$, it takes a remarkably simple form
$x^{-1}-x=N_f$.} It reads 
$T_c=\frac{g^2}{8\pi}\left(\sqrt{N_f^2+4}-N_f\right)$
and vanishes monotonically with the increase of $N_f$
(i.e. with the increase of the strength of the monopole--anti-monopole interaction in the molecule) from 
$\left.T_c\right|_{N_f=0}=g^2/(4\pi)$ to $T_c\stackrel{N_f\to\infty}{\longrightarrow}0$. 
The critical number of flavors can be estimated from the condition
$T_c\sim\zeta^{1/3}$ at large $N_f$. This yields an exponentially large number, namely
$N_f\sim\frac{(g^2/M_W)^{7/6}}{4\pi\delta^{1/3}}
\exp\left(\frac{4\pi\epsilon M_W}{3g^2}\right)$. 
Instead, requiring for both fugacities to be of the order of $\Lambda^2$ at the critical temperature, 
we obtain from the equation 
\begin{equation}
\label{nf}
-S_0+\left(2-\Delta_{N_f}\right)\lambda=-\beta M_W+(2-\tilde\Delta)\lambda=0
\end{equation}
the following formula: $T_c=\frac{g^2}{4\pi}\frac{2+\epsilon-N_f}{1+2\epsilon}$.~\footnote{
As should be, at $N_f=0$, this expression reproduces the above-cited result of Ref.~\cite{sk}. The above-discussed formal limits 
$\epsilon\to 0$ (which now corresponds to 3d compact QED with massless quarks~\cite{q}) and $\epsilon\to\infty$ are, of course, 
reproduced correctly as well.} We see that, in the general case 
outside the BPS limit, the critical number of flavors stemming from this formula is only 4, rather than an exponentially 
large number following from the condition $\Delta_{N_f}=\tilde\Delta$.

The BPS limit requires a special study. As we will see in a moment, naively taking the limit 
$\epsilon\to 1$ in the obtained formula for $T_c$, one gets the wrong expression $T_c=\frac{g^2}{12\pi}(3-N_f)$. It however yields the 
same critical number of flavors, 3, as the correct expression for $T_c$. To derive the latter, one should take into account that, 
in the BPS limit, the scaling dimension of monopoles reads $\Delta=8\pi T/g^2$ (rather than $4\pi T/g^2$ as everywhere outside 
this limit). That is because, in this limit ($M_H=0$), 
the monopole--anti-monopole interaction in a molecule through the scalar part of the 
't~Hooft--Polyakov monopole solution, Eq.~(\ref{scal}), 
becomes as important as their ordinary interaction through the vector part, Eq.~(\ref{vect}). 
Formally, this means that,
in the Coulomb potential of a monopole--anti-monopole pair, one should replace $q_aq_b$ by $q_aq_b-1$, where $q_a, q_b=\pm 1$ are the 
charges of a monopole and an anti-monopole in the units of $g_m$ (see Ref.~\cite{dc} for a review of other effects produced by the 
propagating Higgs field).~\footnote{Due to this fact, the interaction between a monopole and an anti-monopole doubles, whereas 
the interactions monopole--monopole and anti-monopole--anti-monopole vanish. In particular, in the continuum limit of 3d 
compact QED, based on such monopole--anti-monopole ensemble, the inverse Berezinsky--Kosterlitz--Thouless phase transition
(from the plasma to the molecular phase) occurs at the temperature $g^2/(4\pi)$ (rather than $g^2/(2\pi)$~\cite{gg} as outside 
the BPS limit).} Substituting this value of $\Delta$ into Eq.~(\ref{nf}) (where $\epsilon$ should be set equal to 1), 
we obtain for the critical temperature $T_c=\frac{g^2}{16\pi}(3-N_f)$. We see that, 
as stated above, the critical number of flavors in this case is indeed 3, although the expression for the critical temperature is 
different from $\frac{g^2}{12\pi}(3-N_f)$. Notice that, 
even in the absence of fermions, the critical temperature in the BPS limit is $\frac{3g^2}{16\pi}$ and not
$\frac{g^2}{4\pi}$, as could have been naively expected.~\footnote{These results can be compared with those one gets from just 
equating $\Delta_{N_f}$ in the BPS limit with $\tilde\Delta$. This yields the equation $2g_m^2T+4\pi N_f=g^2\beta$, whose solution 
is $T_c=\frac{g^2}{16\pi}\left(\sqrt{N_f^2+8}-N_f\right)$, that leads again to an exponentially large critical number of flavors.
In particular, in the absence of quarks, the critical temperature stemming from this formula is $\frac{g^2}{4\pi\sqrt{2}}$. As we have argued,
these results are as erroneous as their counterparts outside the BPS limit.}

Finally, since monopoles are instantons in 3d, 
it is worth comparing these results with those of the instanton liquid in 4d QCD, where the same effect of binding
of instantons into molecules due to massless fundamental flavors has been found~\cite{ilgsh}. In particular,
the critical number of flavors (at which the temperature of the chiral phase transition reaches zero) has been shown to be
around 5~\cite{schsh}. The order of the 
transition is second for 2 massless flavors and first from 3 flavors on.~\footnote{In particular, the phase transition never occurs to be of the 
Berezinsky--Kosterlitz--Thouless type, as in the continuum limit of 3d compact QED.} 
In the GG model,
where the phase transition is associated with the restoration of the 
so-called magnetic $Z_N$-symmetry, its order  
in the SU($N$)-case is the same as in the $Z_N$-invariant 2d spin models~\cite{gg1, ggrev}. However, due to the completely different mechanisms of 
confinement in the 3d GG model and 4d QCD,~\footnote{While in the 3d GG model confinement 
is produced by magnetic monopoles already at weak
coupling, in 4d QCD it is argued to be due to some stochastic background fields and holds only at strong coupling.} one should not expect any
precise correspondence between the results obtained in these two theories.
For instance, the difference is reflected already in the fact that the temperature of dimensional reduction in the 3d GG model is
exponentially smaller than the deconfinement critical temperature, whereas, in 4d QCD, it is approximately twice larger.

\section{Conclusions}
In the first part of this Letter, we have explored the large-distance quark--anti-quark potential in the GG model. 
The fact that the quark--anti-quark string is not just a free Nambu--Goto one, but passes through W-bosons, yields
positive corrections to the fundamental-string tension and to the coefficient at the $1/R$-term in the potential corresponding 
to the Nambu--Goto string in 3d [Eqs.~(\ref{spectrum}) and (\ref{coeffs})]. The known statement that W-bosons 
are irrelevant due to their large mass is proven in the sense that the relative correction to the 
string tension is exponentially small. Rather, the coefficient 
at the $1/R$-term in the quark--anti-quark potential increases, with respect to the value it has in the Nambu--Goto-string model, 
by a significant factor $4\sqrt{2}$.
It is natural to ask whether such a piecewise string with heavy adjoint nodes appears somewhere in QCD. One situation of this kind 
is realized within QCD in the so-called 
maximal Abelian gauge, where off-diagonal gluons acquire a large mass, about 1.2 GeV~\cite{offdiag}. Another situation where 
we meet this sort of string is QCD above the deconfinement critical temperature but below the temperature of dimensional reduction.
There, the role of heavy adjoint nodes is played by the $A_0$-gluons, whose mass at the one-loop level reads (see e.g. Ref.~\cite{ay})
$\sqrt{\frac{N_c}{3}+\frac{N_f}{6}}gT+{\cal O}(g^2T)$. The spectrum of such a piecewise string at finite temperatures will be studied
in a separate publication.

In the second part of the Letter, we considered the finite-temperature GG model in the presence of $N_f$ massless fundamental quarks.
An additional attraction of a quark and an anti-quark inside a molecule, produced by quark zero modes in the monopole field,
is known, at the example of 3d compact QED, to decrease the value of $T_c$. An interesting issue is the critical number of flavors, at which
the deconfining phase transition occurs at the same exponentially small (with respect to $T_c$) temperatures as the dimensional reduction. 
The results for this number are as follows: 2 for 3d compact QED~\cite{q}, 3 for the GG model in the BPS limit, 4 -- outside this limit.
The latter two numbers, as well as the formula for $T_c$ as a function of $N_f$ (both in and out of the BPS limit) were obtained in Section~3.
In particular, the number 4, obtained in the general case outside the BPS limit, is remarkably close to 5 -- the critical number of flavors,
at which the instanton-liquid model of 4d QCD passes to the molecular phase (that leads to the chiral phase transition)
at vanishingly small temperatures. 
Note finally, that this interesting similarity can also be viewed from the other side. 
Indeed, the GG model with quarks is nothing but 3d QCD with an additional
adjoint Higgs field, where just instantons (=monopoles) are different from those of 4d QCD.

\section*{Acknowledgments}
I am grateful to A.~Di~Giacomo, D.~Ebert, H.~J.~Pirner, E.~V.~Shuryak,  Yu.~A.~Simonov, and D.~T.~Son 
for stimulating discussions. I would also like to thank 
the Alexander~von~Humboldt foundation for the financial support and 
the staff of the Institute of Physics of the Humboldt University of 
Berlin for the cordial hospitality.

\section*{Appendix}
Let us consider a more general Hamiltonian, namely 
$H=\sum\limits_{n}^{}\left[\frac{p_n^2}{2m_n}+
\frac{K}{2}(x_{n+1}-x_n)^2\right]$, 
where $m_n=m$ for $n$ even and $m_n=M$ for $n$ odd.
The Hamilton equations 
yield $m_n\ddot x_n=K(x_{n+1}+x_{n-1}-2x_n)$ or, separately for 
$y_n\equiv x_{2k+1}$ and $z_n\equiv x_{2k}$,

$$
M\ddot y_n=K(z_{n+1}+z_n-2y_n),~~ 
m\ddot z_n=K(y_n+y_{n-1}-2z_n).$$
Seeking solutions in the form of plane waves, $y_n={\rm e}^{i(qn-\omega t)}y_q$, 
$z_n={\rm e}^{i(qn-\omega t)}z_q$, where $q$ is the momentum lying in the 
first Brillouin zone, $-\pi<q<\pi$, one obtains

$$
M\omega^2y_q=K\left(2y_q-(1+{\rm e}^{iq})z_q\right),~~
m\omega^2z_q=K\left(2z_q-(1+{\rm e}^{-iq})y_q\right).$$
The dispersion law is therefore determined by the characteristic equation

$$
\det\left(
\begin{array}{cc}
M\omega^2-2K& K(1+{\rm e}^{iq})\\
-K(1+{\rm e}^{-iq})& 2K-m\omega^2
\end{array}
\right)=0,$$
or $\omega^4-\frac{2K}{\mu}\omega^2+\frac{4K^2}{Mm}\sin^2\frac{q}{2}=0$, where 
$\mu=\frac{Mm}{M+m}$. This yields 

$$\omega_\pm^2(q)=\frac{K}{\mu}\left(1\pm\sqrt{1-\frac{4\mu^2}{Mm}\sin^2\frac{q}{2}}\right),$$
where ``$+$'' corresponds to the optical mode, while ``$-$'' to the acoustic one. In the 
particular case $M=m=M_W$ under study, we have 

$$
\omega_{+}^2(q)=\frac{4K}{M_W}\cos^2\frac{q}{4},~~
\omega_{-}^2(q)=\frac{4K}{M_W}\sin^2\frac{q}{4}.$$
The energy of the quark-antiquark string, thus, reads

$$
E(R)=\sigma R+2\sqrt{\frac{\sigma}{M_Wl}}\sum\limits_{q}^{}\left(\left|\sin\frac{q}{4}\right|+
\left|\cos\frac{q}{4}\right|\right).$$ 
Using finally the formula

$$
\sum\limits_{n=1}^{N}\sin\frac{\pi n}{2(N+1)}=\sum\limits_{n=1}^{N}\cos\frac{\pi n}{2(N+1)}=
\frac{1}{\sqrt{2}}\frac{\sin\frac{\pi N}{4(N+1)}}{\sin\frac{\pi}{4(N+1)}},$$
we arrive at Eq.~(\ref{spectrum}).


\end{document}